\documentclass{article}
\usepackage{spconf,amsmath,graphicx,hyperref}

\usepackage{amsmath,amssymb,amsfonts}
\usepackage{algorithmic}
\usepackage{graphicx}
\usepackage{textcomp}
\usepackage{xcolor}
\def\BibTeX{{\rm B\kern-.05em{\sc i\kern-.025em b}\kern-.08em
    T\kern-.1667em\lower.7ex\hbox{E}\kern-.125emX}}

\usepackage{url,times,booktabs, tabularx}
\usepackage{hyperref}
\usepackage{multirow}
\usepackage{lipsum}
\usepackage{enumitem}
\usepackage{multirow}
\usepackage{amssymb}
\usepackage{pifont}
\usepackage{hyperref}
\usepackage{subfig} 
\usepackage{booktabs} 

\usepackage[
  backend=biber,doi=false,isbn=false,
  style=ieee,
  minnames=1,
  maxcitenames=2, maxbibnames=6
]{biblatex}

\title{Speech Content Privacy in Environmental Sound Recordings using Segment-wise Waveform Reversal}
\name{Modan Tailleur$^{1}$\thanks{MT is funded by the AIby4 project (ECN and ANR-20-THIA-0011).}, Mathieu Lagrange$^{1}$, Pierre Aumond$^{2}$, Vincent Tourre$^{3}$}
\address{$^{1}$ Nantes Université, École Centrale Nantes, CNRS, LS2N, UMR 6004, Nantes, France\\
$^{2}$ Université Gustave Eiffel, CEREMA, UMRAE, Bouguenais, France\\
$^{3}$ Nantes Université, ENSA Nantes, École Centrale Nantes, CNRS, AAU-CRENAU, Nantes, France}

\begin{document}
%
\maketitle
\begin{abstract}
Environmental sound recordings often contain intelligible speech, raising privacy concerns that limit the analysis, sharing, and reuse of the data. In this paper, we introduce a method that renders speech unintelligible while preserving both the integrity of the acoustic scene and the overall audio quality. Our approach involves reversing waveform segments to distort speech content identified using a voice activity detection and speech separation pipeline in order to render the speech unintelligible.

We consider a three-part evaluation protocol that assesses: 1) speech intelligibility using Word Error Rate (WER), 2) sound sources detectability using Sound source Classification Accuracy-Drop (SCAD), and 3) audio quality using the Fréchet Audio Distance (FAD). 
Compared to two other state-of-the-art systems, our method is the only one to achieves satisfactory speech intelligibility reduction (97.9\% WER), with minimal degradation of the sound sources detectability (2.7\% SCAD), and good perceptual quality. An ablation study further highlights the contribution of each component of the speech content privacy enforcement pipeline.  

\end{abstract}
\begin{keywords}
Speech Content Privacy Enforcement, Field Audio Recording, Environmental Audio
\end{keywords}
\section{Introduction}
\label{sec:intro}

Field recording of environmental audio often involves deploying acoustic sensors in public or semi-public spaces to passively capture audio data \cite{bello_sonyc_2019, mietlicki_innovative_2015, ooi_strongly-labelled_2021, vidana2020low, mydlarz2017implementation, arce2021fiware, fallis2020power, peng2024environmental, abesser2021idmt}. However, these recordings frequently contain human speech, raising privacy concerns. As a result, researchers in environmental audio face the dual challenge of protecting speech privacy while preserving the rich contextual information embedded in the recordings.  In many cases, preserving the integrity of the background environment is thus essential for accurate interpretation of the acoustic environment. When it comes to voice, in certain cases, it may be enough to reduce intelligibility by altering the speech content. In more sensitive applications, vocal characteristics—such as emotional tone or speaker identity—can also compromise privacy. At the same time, these features may carry valuable cues, for example about gender or age, and can be important to retain depending on the goals of the analysis. In urban soundscapes, for instance, speech is often one of the dominant sound sources and conveys meanings and emotions that are closely tied to its high-level acoustic properties. Simply removing or heavily modifying speech in such contexts would not only eliminate these important cues but also potentially distort the relative level of vocal sounds with respect to the background environment.

In this paper, we focus on an application scenario that requires enforcing speech content privacy while preserving both the structural integrity of the acoustic scene and the perceptual quality of the audio. We target use cases where recordings may be shared internally for tasks like: annotation, listening-based analysis, quality control, or collaborative research. In such cases, the risk of intentional attempts to recover clean speech for potentially harmful uses is negligible, but there remains a need to ensure that intelligible speech is not accessible, in order to protect the privacy of individuals recorded and allow the listener to focus her/his listening on the quality of the sound environment.

\begin{figure*}[!htb]
\begin{center}
\includegraphics[width=\linewidth]{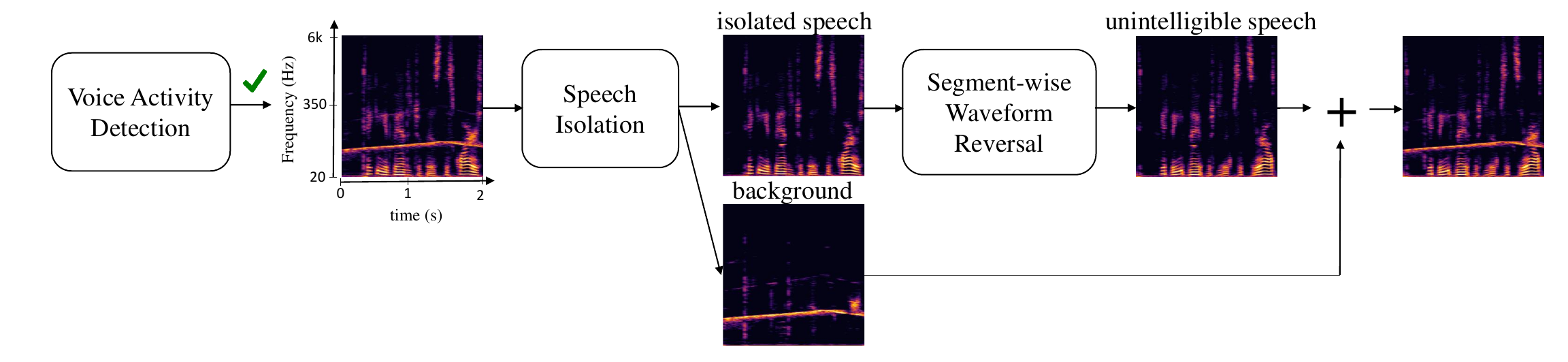}
\end{center}
\vspace{-0.4cm}
\caption{Pipeline for speech content privacy enforcement. If speech is detected in a given segment, the speech is isolated, shuffled and mixed back with the remaining background.}
\vspace{-0.4cm}
\label{fig:pipeline}
\end{figure*}

To this aim, we propose an efficient method to enforce speech content privacy based on Segment-wise Waveform Reversal (SWR), which involves segmenting the audio in regions of low energy and reversing the waveform within each segment. To enhance both audio quality and sound sources detectability, our approach applies SWR only on isolated speech by considering a Voice Activity Detection (VAD) module and a source separation module to isolate the speech component, as shown on Figure \ref{fig:pipeline}. Source code and audio examples are available\footnote{Companion page: \url{https://modantailleur.github.io/paperSpeechContentPrivacyEnforcement/}}



\section{Previous work}
\label{sec:previous_work}

Burkhardt et al. \cite{burkhardt_masking_2023} introduced a speech content anonymization method designed for voice only recordings.  Their method is based on {Fragmentation} and {Reordering}. The fragmentation stage divides the audio into short segments by identifying low-energy regions that are less likely to introduce perceptual artifacts. Segment boundaries are thus chosen at the nearest zero-crossing to the point of lowest Root Mean Square (RMS) energy within a region of interest (ROI). This strategy allows cuts to be made in low-energy regions, thereby minimizing audible artifacts such as clicks or pops in the subsequent processing. In the reordering stage, the resulting fragments are randomly shuffled such that no two consecutive segments from the original sequence are reconnected—unless no other reordering is possible. 

To enforce both speech content privacy and speaker identity protection in more complex audio mixtures, Cohen-Hadria et al. \cite{cohen-hadria_voice_2019} proposed a two-stage pipeline. First, they applied source separation using a U-Net trained on synthetic mixtures of environmental sounds and Librispeech \cite{panayotov_librispeech_2015} utterances. Then, the separated speech channel is anonymized using either low-pass filtering or Mel-Frequency Cepstral Coefficents (MFCC) inversion. The MFCC inversion method, which yielded more consistent results, involves extracting the first five MFCCs, applying inverse discrete cosine transform followed by decibel scaling to obtain a coarse mel spectrogram, converting it to a linear-frequency power spectrogram via a non-negative least squares (NNLS) solver, estimating phase with the Griffin-Lim algorithm, and finally reconstructing the waveform with inverse short-time Fourier transform.

Our goal differs slightly from these two approaches. Unlike Burkhardt et al. \cite{burkhardt_masking_2023}, we focus on complex acoustic scenes that include both speech and non-speech events, rather than isolated voice recordings. While Cohen-Hadria et al. \cite{cohen-hadria_voice_2019} also enforced speech content privacy on complex audio mixtures, their approach additionally aimed at speaker anonymization, while we aim at preserving the timbre of the speech content. 

\section{Method}
\label{sec:methods}

We propose a 4-step approach: 1) detect speech-containing audio segments using voice activity detection (VAD), 2) apply speech isolation to isolate speech from the background, 3) reduce speech intelligibility using our SWR method, 4) mix the modified speech back with the residual background sound. The overall pipeline is illustrated in Figure \ref{fig:pipeline}.

\subsection{Voice Activity Detection (VAD) \label{sec:vad}}

We apply Voice Activity Detection (VAD) to determine the presence of speech using  BEATs \cite{chen_beats_2022} pre-trained classification algorithm. BEATs is an attention-based audio classification model trained on AudioSet \cite{gemmeke_audio_2017}, comprising approximately 86 million parameters. It supports multi-label classification over 527 classes from the AudioSet ontology, covering a broad range of sound sources. Among these output classes, the model includes a "Speech" category, which we use for VAD. We apply a threshold of 0.3 to the speech class logits, favoring a more inclusive detection strategy to better align with privacy-preserving goals. This threshold was selected via a grid search. Voice activity detection is performed on non-overlapping 500ms windows with 440ms hop length.

\subsection{Speech Isolation (SI)}

Where voice activity is detected, we apply a pre-trained Speech Isolation (SI) algorithm to isolate the speech component. We use the Hybrid Demucs (HDemucs) model \cite{defossez_hybrid_2022}, trained on the MUSDB-HQ dataset \cite{musdb18-hq}. 
HDemucs is trained on high-resolution music data, which allows it to handle full-bandwidth signals at a high sampling rate of 44.1 kHz.

\subsection{Intelligibility Reduction}

On the separated speech, we apply a deterministic method based on a Segment-Wise Waveform Reversal (SWR) algorithm. We identify low-energy regions based on RMS energy calculations to segment the audio, as in  Burkhardt et al. \cite{burkhardt_masking_2023}. We use a threshold of -6dBFS for identifying low energy regions. This process is applied per texture frames of two seconds. For each segment obtained during fragmentation, we then apply waveform reversal, which consists of reversing the audio samples within each segment. We use an overlap-add technique with a 5\% overlap applied at both the segment and texture frame levels to reduce artifacts introduced by discontinuities at reconnected waveform boundaries.


\section{Dataset}
\label{sec:format}

Our evaluation is based on a simulated dataset designed to reflect realistic sound environment conditions. It consists of speech mixed with diverse background sounds, as well as voice-free audio excerpts. To build this dataset, we combine speech samples from LibriSpeech \cite{panayotov_librispeech_2015} with environmental recordings from SONYC-UST \cite{cartwright_sonyc-ust-v2_2020}. LibriSpeech \cite{panayotov_librispeech_2015} is a large-scale corpus of English read speech, designed for training and evaluating automatic speech recognition (ASR) systems. 
SONYC-UST-V2 \cite{cartwright_sonyc-ust-v2_2020} is a large-scale, multi-label dataset for urban sound tagging. It comprises 18,510 ten-second audio clips, sampled at 48 kHz, recorded between 2016 and 2019 by a network of over 50 acoustic sensors deployed throughout New York City as part of the Sounds of New York City (SONYC) \cite{bello_sonyc_2019} project. 

To construct our dataset from SONYC-UST and LibriSpeech, we begin by selecting from the SONYC-UST evaluation set all audio samples annotated with exactly one label among the following sound sources: engine, jackhammer, chainsaw, car horn, siren, music, and dog. 
We then balance this subset by randomly sampling an equal number of recordings per class, resulting in a total of 742 samples.

Next, we randomly shuffle the LibriSpeech evaluation set and select 371 speech recordings with durations closest to 10 seconds. These are then mixed with 371 SONYC-UST recordings, after downsampling both audio sources to a common 44.1 kHz sampling rate. Prior to mixing, we normalize both signals to the same root mean square energy (RMSE) and apply a 6 dB gain to the SONYC-UST signal to simulate realistic background noise conditions. Finally, the resulting mixture is normalized to its maximum amplitude.

As a result, we obtain an evaluation dataset of 742 audio mixtures of 10 seconds each: 371 containing only voice-free environmental sounds and 371 containing speech over background environmental sound. The dataset, named \textit{CitySpeechMix}, available online \footnote{\url{https://zenodo.org/records/15405950}}, is approximately 2 hours long. Note that this dataset is used solely for evaluation and is not employed to train any of the algorithms.

\begin{figure}[t!]
\centering
\begin{tabular}{l|r|r|r}
\textbf{Method} & \multicolumn{1}{c|}{\textit{WER \%~$\uparrow$}} & \multicolumn{1}{c|}{\textit{SCAD \%~$\downarrow$}} & \multicolumn{1}{c}{\textit{FAD~$\downarrow$}} \\
\midrule
Original audio & 9.4{ ±0.6} & 0.0{ ±3.1} & 0.0 \\
White noise & 98.5{ ±0.1} & 41.0{ ±3.8} & 140.9 \\
\midrule
Cohen-Adria \cite{cohen-hadria_voice_2019} & 83.6{ ±1.0} & 25.1{ ±3.9} & 21.8 \\
Burkhardt \cite{burkhardt_masking_2023} & 86.7{ ±0.9} & \textbf{2.7}{ ±3.2} & 3.1 \\
Ours & \textbf{97.9}{ ±0.2} & \textbf{2.7}{ ±3.2} & \textbf{1.4} \\
\midrule
- w SepFormer & \textbf{97.9}{ ±0.2} & \textbf{1.0}{ ±3.1} & 2.7 \\
\midrule
- w/o VAD & \textbf{98.1}{ ±0.2} & 4.1{ ±3.3} & 2.7 \\
- w/o SI & \textbf{98.0}{ ±0.2} & \textbf{1.3}{ ±3.1} & 2.4
\end{tabular}
\caption{Evaluation of speech content privacy enforcement, addressing 1) intelligibility reduction, evaluated using WER, 2) sound scene preservation, evaluated using SCAD, and audio quality, evaluated using FAD. Bold highlights  best performance and results with variance estimates not statistically different from the best-performing method (paired t-test for WER, McNemar test for SCAD, $p > 0.01$, 95\% confidence intervals).  
\vspace{-0.4cm}
}
\label{tab:results_combined}
\end{figure}

\section{Metrics}


\subsection{Intelligibility Reduction}

To evaluate how effectively the algorithm reduces speech intelligibility, we compute the Word Error Rate (WER) using several state-of-the-art automatic speech recognition (ASR) models. WER measures the discrepancy between reference and predicted transcriptions, expressed as a percentage, with lower values indicating better transcription accuracy. To prevent extreme misrecognition cases from skewing the results, we cap the WER at 100\%. We compare the reference LibriSpeech transcriptions to the outputs of the following ASR models: Wav2Vec2 fine-tuned for ASR 
\cite{baevski_wav2vec_2020}, the Whisper large-v3 model \cite{radford_robust_2023}, Fairseq S2T \cite{wang_fairseq_2020}, and the CRDNN model from SpeechBrain \cite{ravanelli_speechbrain_2021}, specifically the \textit{asr-crdnn-rnnlm-librispeech} pipeline. We then compute the average WER across all models to obtain an overall measure of intelligibility reduction.

\subsection{Sound Sources Detectability }

To assess the impact of the anonymization method on the acoustic scene, we evaluate whether the sound sources remain as recognizable as in the original recordings. Specifically, we measure whether anonymization degrades the performance of a sound source classification model. We use the BEATs model, previously introduced in Section \ref{sec:vad}. Our analysis focuses on the eight sound classes of our dataset: speech, engine, jackhammer, chainsaw, car horn, siren, music, and dog. 

For the 371 mixed clips, each audio contains two sources: speech and one background class. A source is thus considered detected if its logit is ranked in the top 2 of the 8 classes. For the voice-free audio excerpts (SONYC-UST only), the source must rank top 1 to be considered detected, as there is only one class in the audio. We report accuracy on both original and anonymized audio. The difference between them, referred to as the Source Classification Accuracy-Drop (SCAD), reflects the degradation in source detection due to the privacy enforcement. A larger drop indicates greater impact, thus lower preservation of the acoustic scene.

\subsection{Audio Quality}




The Fr\'echet Audio Distance (FAD) \cite{kilgour_frechet_2019} has been shown to correlate well with perceived audio quality across multiple studies \cite{tailleur_correlation_2024, choi_foley_2023, gui_adapting_2024}, considering that the reference dataset is of good quality. It has been proposed as an adaptation of the Fr\'echet Inception Distance (FID) \cite{heusel_gans_2017} for audio quality assessment. FID and FAD compare the distribution of two datasets in a given embedding space. Based on the findings of \cite{tailleur_correlation_2024}, we use the PANN-Wavegram-Logmel model \cite{kong_panns_2020} to extract audio embeddings, as it has been shown to correlate better with perceptual ratings than the originally proposed VGGish model \cite{hershey_cnn_2017}. 

\section{Evaluation}
\label{sec:results}

\subsection{Reference Methods}

We implement Burkhardt et al. anonymization method described in section \ref{sec:previous_work} on the entirety of each audio, using a threshold of $-6\,\mathrm{dB}$ relative to the maximum level for identifying low energy regions. 

We implement Cohen-Hadria’s method described in section \ref{sec:previous_work}. For the SI step in Cohen-Hadria’s pipeline, we substitute their U-Net model with our HDemucs-based SI, due to the unavailability of the original implementation. 

Additionally, we compare our HDemucs-based SI approach with a pre-trained SepFormer model \cite{subakan2021attention}, an attention-based architecture provided by the SpeechBrain toolkit \cite{ravanelli_speechbrain_2021}. This SepFormer is trained on the WHAM! dataset \cite{wichern2019wham}, which contains mixtures of speech and background environmental audio. It operates at a 16 kHz sampling rate. Finally, we include a white noise baseline where the entire audio is replaced with white noise.


\subsection{Results}

The evaluation results of the different anonymization methods are presented in Table~\ref{tab:results_combined}. Our method outperforms or matches all reference methods across the three evaluation metrics. It achieves a WER of 97.9\%, which is close to the performance obtained with white noise generation, indicating that speech is indeed unintelligible. It results in an accuracy drop of only 2.7\%, demonstrating strong preservation of the acoustic scene. Additionally, our method yields the lowest FAD among all the evaluated approaches, indicating the best preservation of overall audio quality.

Replacing the chosen separation module for the Brain SepFormer does not lead to a significant change in terms of WER and SCAD. The resulting audio quality is lower, probably due to the fact that the Brain SepFormer operates at only 16 kHz.

\subsection{Ablation study}

To assess the contribution of each key components of the pipeline, we perform an ablation study of the two main preprocessing components: VAD and SI. As expected, we find that intelligibility reduction is largely unaffected by the removal of individual components, with WER remaining consistently above 97.9\%. For sound sources detectability, the VAD component has the most noticeable impact. Removing it increases SCAD by around 1.5\%, showing that accurate detection of speech is important for preserving non-speech content. The removal of the SI component has no statistically significant effect. The FAD score increases when VAD, or SI are removed, indicating that those two components have an impact on audio quality.

\section{Discussion}
\label{sec:conclusion}

Our proposed speech content privacy enforcement method aims to render speech unintelligible while preserving the surrounding acoustic scene and overall audio quality. The approach is based on Segment-Wise Waveform Reversal, and is further enhanced through a Voice Activity Detection (VAD) stage followed by a Speech Isolation (SI) stage.

Evaluated on a dataset combining speech with urban environmental sounds, the method achieves a Word Error Rate (WER) of 97.9\%, effectively eliminating intelligible speech. It preserves the integrity of the original environment, with only a 2.7\% drop in sound source classification accuracy, and offers the best perceptual quality among compared methods, as reflected by a Fréchet Audio Distance (FAD) of 1.40.

To further improve robustness against privacy adversarial attacks, a segment reordering stage such as the one proposed in \cite{burkhardt_masking_2023} can be added. Indeed, our method relies on deterministic segmentation and simple waveform inversion. As a result, an attacker with knowledge of the segmentation procedure can attempt to recover the original speech content for example by simply reapplying the pipeline to an already-processed signal. 

Using the same two-second texture frame as in the waveform reversal step leads, reordering segments lead to a reduction in audio quality (FAD increases from 1.4 to 1.54). While the difference may appear modest, informal listening indicates a noticeable effect (audio examples are provided in the companion page). This loss in audio quality leads to a substantial gain in robustness. Specifically, applying the proposed pipeline twice without reordering results in a WER of 54.9\%, whereas applying it twice with reordering increases the WER to 93.2\%. These findings emphasize that the reordering step significantly improves the robustness of the speech content privacy enforcement pipeline, at the cost of reducing audio quality. In terms of sound sources detectability, SCAD does not show a statistically significant difference, based on a McNemar test with a p-value threshold of 0.01, suggesting that the acoustic scene remains largely unchanged between the two methods.

The proposed method achieved satisfactory results for the use case considered: internal listening of the recorded audio without display of spoken information. Future work will focus on improving the versatility and robustness of the method to support a wider range of use cases with stricter privacy constraints.




\newpage

\printbibliography

\end{document}